\documentclass[preprint,aps,superscriptaddress,showpacs,preprintnumbers,amsmath,amssymb,floatfix,12pt,nofootinbib]{revtex4}

\usepackage{graphicx}
\usepackage{epsfig}
\newcommand{\bea}{\begin{eqnarray}}
\newcommand{\eea}{\end{eqnarray}}
\def\MeV{\mathop{\rm MeV}\nolimits}
\def\GeV{\mathop{\rm GeV}\nolimits}
\def\Str{\textrm{Str}}
\def\CF{{\cal F}}
\def\CL{{\cal L}}
\def\CM{{\cal M}}
\def\CO{{\cal O}}
\def\CV{{\cal V}}
\def\etal{{\it et al}.\ }

\begin{document}
\bibliographystyle{apsrev}

\title{{\large\bf  $B \rightarrow D^* l\nu$ and $B \rightarrow D
l\nu$ form factors in staggered chiral perturbation theory}}

\author{Jack~Laiho}
\affiliation{Theoretical Physics Department, Fermilab, Batavia, IL
60510}
\bigskip

\author{Ruth~S.~Van~de~Water}
\affiliation{Theoretical Physics Department, Fermilab, Batavia, IL
60510}
\bigskip

\bibliographystyle{apsrev}

\pacs{11.15.Ha, 
      11.30.Rd, 
      12.38.Gc  
}

\begin{abstract}
We calculate the $B\to D$ and $B\to D^*$ form factors at zero
recoil in Staggered Chiral Perturbation Theory.  We consider
heavy-light mesons in which only the light ($u$, $d$, or $s$)
quark is staggered; current lattice simulations generally use a
highly improved action such as the Fermilab or NRQCD action for
the heavy ($b$ or $c$) quark. We work to lowest nontrivial order in the heavy
quark expansion and to one-loop in the chiral
expansion. We present results for a partially quenched theory with
three sea quarks in which there are no mass degeneracies (the
``1+1+1" theory) and for a partially quenched theory in which the
$u$ and $d$ sea quark masses are equal (the ``2+1" theory).  We
also present results for full (2+1) QCD, along with a numerical
estimate of the size of staggered discretization errors. Finally,
we calculate the finite volume corrections to the form factors and
estimate their numerical size in current lattice simulations.
\end{abstract}

\maketitle

\section{Introduction}

The CKM matrix element $|V_{cb}\,|$, which places an important
constraint on the apex of the CKM unitarity triangle through the
ratio $|V_{ub}/V_{cb}\,|$, can be determined from experimental
measurements of exclusive semileptonic $B$-meson decays combined
with theoretical input.   Because experiments measure the product
$(\CF(1)\cdot|V_{cb}\,|)^2$, where $\CF(1)$ is the $B\to D$ or
$B\to D^*$ hadronic form factor at zero recoil, the precision of
$|V_{cb}\,|$ is limited by the theoretical uncertainty in
$\CF(1)$.  Although, in principle, both form factors can be
calculated nonperturbatively using lattice QCD, in practice,
direct calculations of the $B\to D$ and $B\to D^*$ hadronic matrix
elements are plagued by large statistical and systematic errors.
Hashimoto \etal therefore proposed a method for calculating
$\CF(1)$ on the lattice using double ratios of matrix elements in
which most of the statistical and systematic errors
cancel~\cite{Hashimoto:1999yp,Hashimoto:2001nb}.  This method
provides the key theoretical ingredient necessary to allow a
precise lattice determination of the $B\to D$ or $B\to D^*$ form
factors, and hence a precise determination of $|V_{cb}\,|$.

Recently $\CF_{B\to D}$ was calculated~using 2+1 flavors of
dynamical staggered quarks~\cite{Okamoto:2004xg}, and the
analogous calculation of $\CF_{B\to D^*}$ will be done in the near
future. Because staggered quarks are computationally cheaper than
other standard fermion discretizations, staggered simulations
offer the lightest dynamical quark masses currently
available~\cite{Kennedy:2004ae}.  This result therefore has a
smaller systematic error associated with chiral extrapolation than
previous quenched results~\cite{Hashimoto:2001nb}.  It is known,
however, that the $\CO(a^2)$ discretization errors associated with
staggered fermions are numerically significant in current lattice
simulations and must also be accounted for in the chiral and
continuum extrapolation of staggered  lattice
data~\cite{Aubin:2004fs}.  This procedure is well-established in
the light meson sector: use of staggered chiral perturbation
theory~\cite{Lee:1999zx,Aubin:2003mg, Aubin:2003uc, Sharpe:2004is}
functional forms for extrapolation of staggered lattice data has
allowed precise determinations of light meson masses, meson decay
constants, and even quark masses~\cite{Aubin:2004fs}.  Staggered
chiral perturbation theory was recently extended to heavy-light
mesons (in which only the light quark is staggered) by Aubin and
Bernard~\cite{Aubin:2005aq}, and has been successfully used in the
extrapolation of the $D$-meson decay constant~\cite{Aubin:2005ar}.

In this paper we use heavy-light staggered chiral perturbation
theory to calculate the $B\to D$ and $B\to D^*$ form factors at
zero recoil.\footnote{Note that a subset of our results was
presented in Ref.~\cite{Laiho:2005np}. }  The resulting functional
forms can then be used to extrapolate staggered lattice data to
the continuum and to the physical pion mass.  Accounting for
staggered discretization errors in this way is essential for a
precise lattice determination of these form factors, and
consequently of $|V_{cb}\,|$.

\bigskip

This paper is organized as follows.  We review staggered chiral
perturbation theory for heavy-light mesons in
Section~\ref{sec:SChPT}. We then calculate the $B\to D$ and $B\to
D^*$ form factors at zero recoil for a 1+1+1 PQ theory, a 2+1 PQ
theory and full (2+1) QCD in Section~\ref{sec:BtoD}.  Next, in
Section~\ref{sec:num}, we plot the $B\to D$ and $B\to D^*$ form
factors using reasonable values for the quark masses and lattice
spacing both with and without the taste-symmetry breaking
contributions. The dramatic change in the shape of the $B\to D^*$
form factor illustrates the necessity of accounting for
taste-breaking in the continuum and chiral extrapolation of
staggered lattice data.  In Section~\ref{sec:FVE} we use the
method of Ref.~\cite{Arndt:2004bg} to calculate the finite volume
corrections to the $B\to D$ and $B\to D^*$ form factors.  We then
estimate the numerical size of these finite volume corrections in
current lattice simulations;  we find them to be very small --
only one part in $10^4$. In Section~\ref{sec:Conc} we conclude.
The Appendix contains additional formulae necessary to understand
our form factor results.  It follows the conventions of
Ref.~\cite{Aubin:2003uc}.

\section{Staggered $\chi$PT  with Heavy-Light Mesons}
\label{sec:SChPT}

In this section we review staggered chiral perturbation theory
(S$\chi$PT) for heavy-light mesons, which was developed in
Ref.~\cite{Aubin:2005aq}.

\bigskip

We first construct the portion of the heavy meson chiral
Lagrangian that only contains light quark fields.  We consider a
partially quenched theory with $n$ flavors of staggered light
quarks.  The detailed construction of the leading-order effective
staggered chiral Lagrangian is given in Ref.~\cite{Aubin:2003mg};
we simply present the results that are necessary for the
calculation of the $B\to D$ and $B\to D^*$ form factors.

We assume that spontaneous symmetry breaking of the $SU(4n)$
chiral symmetry by the vacuum,
\begin{equation}
    SU(4n)_L \times SU(4n)_R \rightarrow SU(4n)_V
\,,
\end{equation}
leads to $16n^2 -1$ pseudo-Goldstone bosons, which we will
generically call pions, that can be collected into an $SU(4n)$
matrix:
\begin{equation}
\Sigma=  \exp(i\Phi / f) \,.
\end{equation}
The matrix, $\Phi$, which contains the pion fields, is traceless
with $4\times 4$ submatrices:
\begin{eqnarray}\label{eq:Phi}
    \Phi & = & \left( \begin{array}{cccc} U & \pi^+ & K^+ & \cdots \\*
    \pi^- & D & K^0 & \cdots \\* K^- & \bar{K^0} & S & \cdots \\*
    \vdots & \vdots & \vdots & \ddots \end{array} \right), \\
    U & = & \sum_{\Xi=1}^{16} U_\Xi T_\Xi \,, \;\; \textrm{etc.}
\end{eqnarray}
where the $SU(4)$ generators,
\begin{equation}\label{eq:T_a}
    T_\Xi = \{ \xi_5, i\xi_{\mu 5}, i\xi_{\mu\nu}, \xi_{\mu},
    \xi_I\}\,,
\end{equation}
are Euclidean gamma matrices and $\xi_I$ is the $4\times 4$
identity matrix. The leading order pion decay constant, $f$, is
approximately $131 \MeV$.  Like the pion matrix, the quark mass
matrix is also $4n \times 4n$, but it has trivial taste structure:
\begin{eqnarray}
    \CM = \left( \begin{array}{cccc} m_u I & 0 &0 & \cdots \\* 0 &
    m_d I & 0 & \cdots \\* 0 & 0 & m_s I & \cdots\\* \vdots &
    \vdots & \vdots & \ddots \end{array} \right)\,.
\end{eqnarray}

Under chiral symmetry transformations,
\begin{eqnarray}
    &\Sigma \rightarrow  L \Sigma R^{\dagger}, \\ & \CM \rightarrow  L \CM R^{\dagger},& \\
    &L \in SU(4n)_L , \;\;\; R \in SU(4n)_R .&
\end{eqnarray}
The standard S$\chi$PT power-counting scheme is:
\begin{equation}
    p_\pi^2/\Lambda_\chi^2 \approx m_q/\Lambda_\textrm{QCD} \approx a^2 \Lambda_\textrm{QCD}^2 \,,
\end{equation}
so the lowest-order, $\CO(p_\pi^2, m_q, a^2)$, staggered chiral
Lagrangian is\footnote{Although we are interested in describing a
Euclidean lattice theory, we choose to perform the calculation in
Minkowski space in order to make intermediate steps comparable to
the continuum literature.  Our results for the form factors will
be independent of this choice.}
\begin{eqnarray} {\cal L}_{\textrm{S$\chi$PT}}^{\textrm{LO}} &=& \frac{f^{2}}{8}
    \Str\big[\partial_{\mu}\Sigma\partial^{\mu}\Sigma\big] +
    \frac{f^{2}\mu}{4}\Str\big[\CM^{\dag}\Sigma+\Sigma^{\dag}\CM\big]
    - \frac{2m^2_0}{3}\big(U_I+D_I+S_I\big)^2 - a^2{\cal V}.
    \nonumber \\ &&
\end{eqnarray}
The staggered potential, $\CV$, splits the tree-level pion masses
into five degenerate groups,
\begin{equation}
    \left(m_{\pi}^2\right)_{\rm LO}=\mu (m_i + m_j) + a^2 \Delta_\Xi \,,
\label{eq:LOmass}\end{equation}
according to their $SO(4)$-taste irrep, $\Xi = I, P, V, A,
T$.\footnote{Note that the splitting $\Delta_P = 0$ because the
taste pseudoscalar pion is an exact lattice Goldstone boson in the
chiral limit.}
  It also leads to hairpin
(quark-disconnected) propagators with multiple poles for
flavor-neutral, taste $V$ and $A$ pions.

\bigskip

We now construct the remaining terms in the heavy meson chiral
Lagrangian. Ref.~\cite{Aubin:2005aq} showed that, at $\CO(a^2)$,
mixed four-fermion operators with both heavy and light quarks
cannot break taste-symmetry.  Because all taste-violation in the
Symanzik action comes strictly from the light quark sector,
discretization errors caused by mixed four-fermion operators can
be categorized as ``heavy-quark errors" and estimated using
standard methods~\cite{Kronfeld:2000ck,Kronfeld:2003sd}.  Thus the
form of the heavy meson portion of the chiral Lagrangian is
identical to that in the continuum, with the exception that the
light quark index can run over both flavor and taste.

Heavy meson chiral perturbation theory (HM$\chi$PT) was first
formulated in Refs.~\cite{Burdman:1992gh,Wise:1992hn} and
generalized to partially quenched QCD in
Ref.~\cite{Sharpe:1995qp}.  Heavy quark spin symmetry allows the
pseudoscalar and vector mesons to be combined into a single field
which annihilates a heavy-light meson:
\begin{eqnarray} H_a=\frac{1+ \not \! v}{2}[\gamma^\mu B^*_{\mu a}+i\gamma_5
B_a],
\end{eqnarray}
where $v$ is the meson's velocity and $a$ labels the flavor and
taste of the light quark within the meson.  Note that, although we
use the letter $B$, the heavy-light meson can be either a $B$, in
which the heavy quark is a $b$, or a $D$, in which the heavy quark
is a $c$.  We also define the conjugate field, $\overline{H}_a
\equiv \gamma_0 H_a^\dag \gamma_0$, which creates a heavy-light
meson.

Under heavy-quark spin symmetry,
\begin{eqnarray}
&H \rightarrow S H, \;\;\; \overline{H} \rightarrow \overline{H}S^{\dagger}& \\
&S \in SU(2),&
\end{eqnarray}
while under chiral symmetry,
\begin{eqnarray}
&H \rightarrow H U^{\dagger}, \;\;\; \overline{H} \rightarrow U \overline{H}& \\
&U \in SU(4n).&
\end{eqnarray}
Interaction terms between heavy-light and pion fields are
constructed using $\sigma = \sqrt{\Sigma} =
\textrm{exp}[i\Phi/2f]$, which is invariant under heavy-quark spin
symmetry but transforms under chiral symmetry as
\begin{eqnarray}
\sigma \rightarrow L \sigma U^{\dagger}=U\sigma R^\dagger, \;\;\;
\sigma^{\dagger} \rightarrow R \sigma^{\dagger} U^{\dagger} = U
\sigma^{\dagger} L^{\dagger}.
\end{eqnarray}
Heavy meson chiral perturbation theory is a joint expansion in the inverse of the
heavy quark mass, $1/m_Q$, and in the residual momentum of the
heavy-light meson, $k$.  Thus the leading order heavy meson
Lagrangian is of $\CO(k)$:
\begin{eqnarray} {\cal L}^{\textrm{LO}}_{\textrm{HM$\chi$PT}} &=& -i \textrm{tr}_D\big[\overline{H}_av^\mu
\big(\delta_{ab}\partial_\mu + iV^{ba}_\mu\big)H_b\big]  +g_\pi
\textrm{tr}_D\big(\overline{H}_a H_b \gamma^\nu \gamma_5 A^{ba}_\nu\big), \label{eq:HMChPT}\end{eqnarray}
\noindent where $V_{\mu} \equiv \frac{i}{2}\big[\sigma^\dag
\partial_\mu\sigma + \sigma\partial_\mu\sigma^\dag\big]$, $A_{\mu}
\equiv \frac{i}{2}\big[\sigma^\dag \partial_\mu\sigma -
\sigma\partial_\mu\sigma^\dag\big]$ and $\textrm{tr}_D$ indicates a trace over Dirac spin indices.
Combining this with the purely pionic terms, the total chiral
Lagrangian for heavy-light mesons in which the light quark is
staggered is
\begin{eqnarray} \CL^{\textrm{LO}} = \CL_{\textrm{HM$\chi$PT}}^{\textrm{LO}} + \CL_{\textrm{S$\chi$PT}}^{\textrm{LO}}. \label{eq:totalL}\end{eqnarray}

\section{Chiral Corrections to $B\to D$ and $B\to D^*$ at Zero Recoil}
\label{sec:BtoD}

The hadronic matrix elements for $B\to D^{(*)}$ depend upon six independent form factors:
\begin{eqnarray}
    \langle D(v')|\overline{c}\gamma^\mu b|\overline{B}(v)\rangle &=& h_{+}(w)(v+v')^\mu + h_{-}(v-v')^\mu , \\
        \langle D^*(v')|\overline{c}\gamma^\mu \gamma_5 b|\overline{B}(v)\rangle &=& -i h_{A_1}(w)(w+1) {{\epsilon}^*}^\mu \nonumber\\
    && + \,i h_{A_2}(w)(v\cdot  {\epsilon}^*)v^\mu + i h_{A_3}(w)(v\cdot  {\epsilon}^*)v'^\mu , \\
    \langle D^*(v')|\overline{c}\gamma^\mu  b|\overline{B}(v)\rangle & = & h_V(w)\epsilon^{\mu\nu\alpha\beta } {{\epsilon}^*}_\nu v'_\alpha v_\beta
\end{eqnarray}
where $w=v \cdot v'$.  In the static heavy quark limit, however,
heavy quark spin symmetry requires that $h_{-} = h_{A_2} = 0$ and
$h_{+}(w) = h_{A_{1,3}}(w) = h_V(w) = \xi(w)$, where $\xi(w)$ is
the universal function for $B\to D^{(*)}$ decays called the
Isgur-Wise function.\footnote{This is true up to radiative
corrections, which only affect $\chi$PT through a modification of
the low energy constants.} Our goal is to calculate the leading
nontrivial contributions to the $B\to D^{(*)}$ form factors at
zero recoil, i.e. when $v'=v$ and $w=1$.

\bigskip

At zero recoil, the hadronic matrix elements depend upon only two
form factors, $h_+(1)$ and $h_{A_1}(1)$:\footnote{Note that the
form factor $h_{-}(1)$ appears in the differential decay rate for
$B\to D$ and is needed in lattice determinations of $|V_{cb}|$
from $B\to D$.  Lattice calculations have shown that this term is
a small correction, and we do not consider here the chiral
corrections to this small quantity.}
\begin{eqnarray} \langle D(v)|\overline{c}\gamma^\mu b|\overline{B}(v)\rangle &=& 2v^\mu h_{+}(1), \label{eq:hplus} \\
     \langle D^*(v)|\overline{c}\gamma^\mu \gamma_5 b|\overline{B}(v)\rangle &=& -i 2{{\epsilon}^*}^\mu h_{A_1}(1). \label{eq:hA1} \end{eqnarray}
In the static heavy quark limit, $\xi(1)$ is normalized to
unity~\cite{Isgur:1989vq}; corrections to this result come from
operators of $\CO(1/m_Q)$.  Operators of $\CO(1/m_Q)$ can be
separated into those that respect heavy-quark spin symmetry and
those that break heavy-quark spin symmetry.  The former cannot
produce logarithmic contributions to $B\to D^{(*)}$ form factors
at leading order in $\chi$PT because they contribute equally to
$B(D)$ and $B^*(D^*)$ masses, so we do not show them here. The
single $\CO(1/m_Q)$ operator that breaks heavy-quark spin symmetry
comes from the interaction between the chromomagnetic moment of
the heavy quark and the light degrees of freedom:
\bea {\delta \cal L} =
\frac{\lambda_2}{m_Q}\textrm{tr}_D[\overline{H}_a \sigma^{\mu
\nu}H_a \sigma_{\mu \nu}]. \label{eq:mcsplit}
\eea
This operator generates a splitting between the $D$ and $D^*$
meson masses, $\Delta^{(c)} = (m_{D^*}-m_D) = -\lambda_2 / 8 m_c$.
It also produces a $B-B^*$ mass splitting, but $\Delta^{(b)}$ is
of $\CO(1/m_b)$ and can be neglected. Finally, we note that there turn out to be no $\CO(1/m_Q)$
corrections to the form factors at zero recoil because of Luke's
theorem \cite{Luke:1990eg}, so the leading nontrivial
contribution to $h_+(1)$ and $h_{A_1}(1)$ is of $\CO(1/m_c^2)$.

In order to calculate the form factors $h_{+}(1)$ and
$h_{A_1}(1)$, we must first map the quark-level $B\rightarrow
D^{(*)}$ operator onto an operator in the chiral effective theory:
\begin{eqnarray}
\bar{c} \gamma^\mu (1-\gamma_5) b \rightarrow -\xi(w) \, \textrm{Str} \big[\bar{H}^{(c)}_{v'} \gamma^\mu (1-\gamma_5) H^{(b)}_{v}\big].
\label{eq:BtoD_Op}\end{eqnarray}
We can then calculate the desired hadronic matrix elements,
Eqs.~(\ref{eq:hplus})--(\ref{eq:hA1}), in the heavy-light meson
effective theory that is described by the Lagrangian in
Eq.~(\ref{eq:totalL}) plus the additional $D-D^*$ mass splitting
term, Eq.~(\ref{eq:mcsplit}).

\begin{figure}
\begin{center}
\includegraphics[scale=0.6]{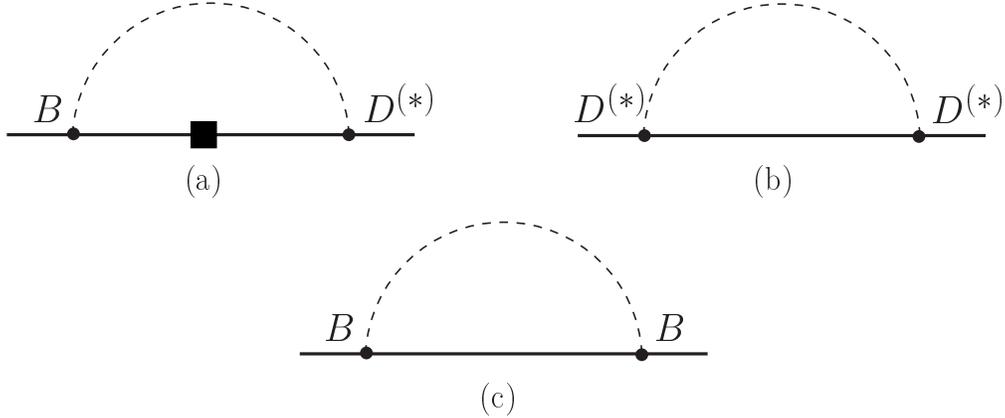}
\end{center}
\caption{One-loop diagrams that contribute to $B\to D^*$.  The
solid line represents a meson containing a heavy quark, and the
dashed line represents light mesons.  The small solid circles are
strong vertices and contribute a factor of $g_\pi$. The large
solid square is a weak interaction vertex.  Diagram (a) is a
vertex correction;  (b) and (c) correspond to wavefunction
renormalization. \label{fig:BtoD_Pion}}
\end{figure}

\begin{figure}
\begin{tabular}{ccc}
    \epsfxsize=2.in \epsffile{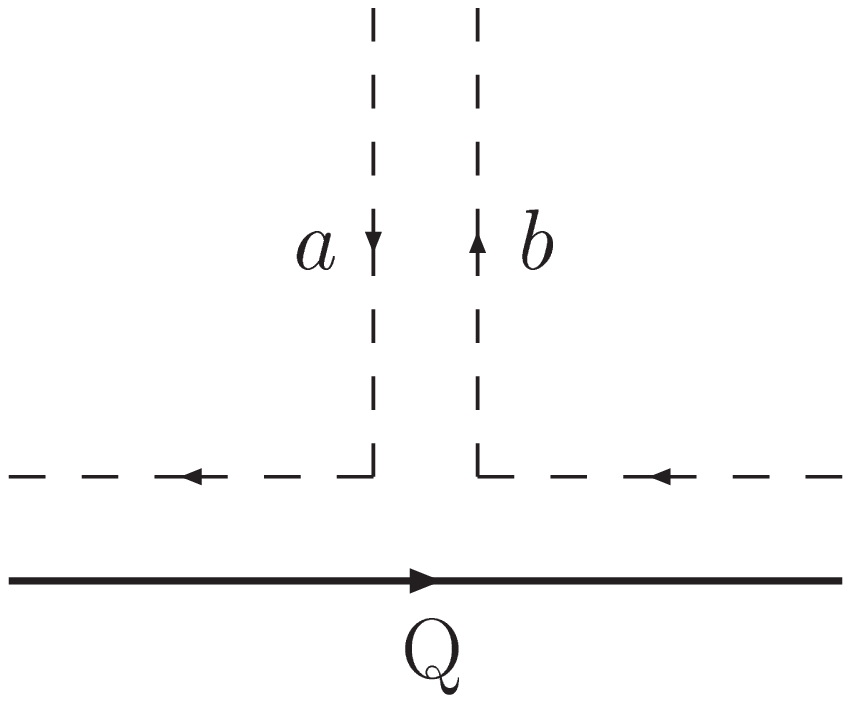} & $\;\;\;\;\;\;\;\;\;\;$ & \epsfxsize=2.in \epsffile{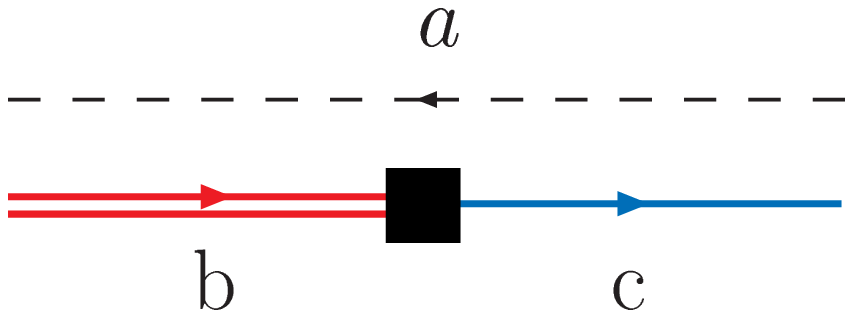} \\
    (a) & $\;\;\;\;\;\;\;\;\;\;$ &  (b)
    \end{tabular}
    \caption{Relevant vertices at the quark level.  Vertex (a) comes from the LO heavy meson chiral
    Lagrangian and is proportional to the coefficient $g_\pi$.  The solid line corresponds to a heavy bottom
    or charm quark while the dashed lines correspond to light staggered quarks of any flavor and taste.
    In this vertex, either one or both heavy-light fields must correspond to a vector meson.
    Vertex (b) comes from the $B\rightarrow D^{(*)}$ operator in Eq.~(\ref{eq:BtoD_Op}).
    The solid double line corresponds to the bottom quark within the $B$- or $B^*$-meson and the solid single line
    corresponds to the charm quark within the $D$- or $D^*$-meson.}\label{fig:vertices}\end{figure}

\begin{figure}
\begin{tabular}{cccccc}
    \epsfxsize=1.75in \epsffile{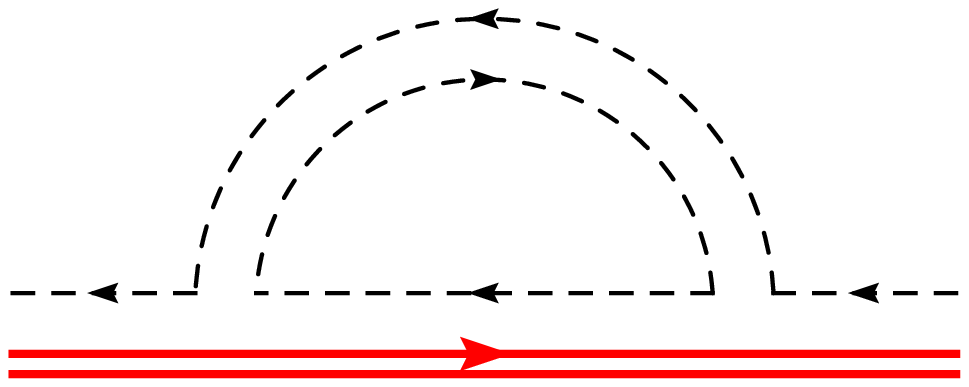} & $\;\;+\;\;$ & \epsfxsize=1.75in \epsffile{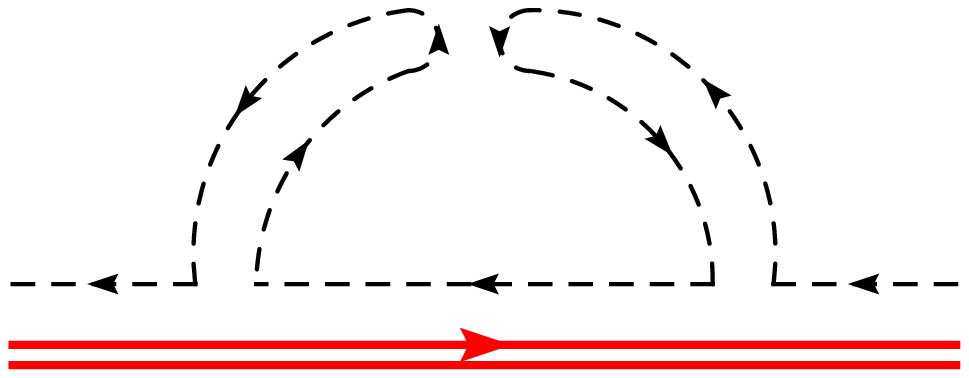} & $\;\;+\;\;$ & \epsfxsize=1.75in \epsffile{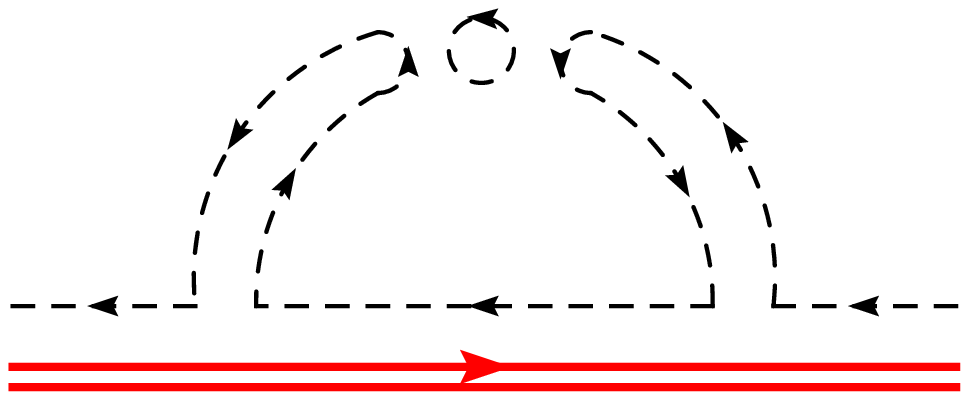} & \;\;\ldots\\
    & & (a) & &  \\
    \epsfxsize=1.75in \epsffile{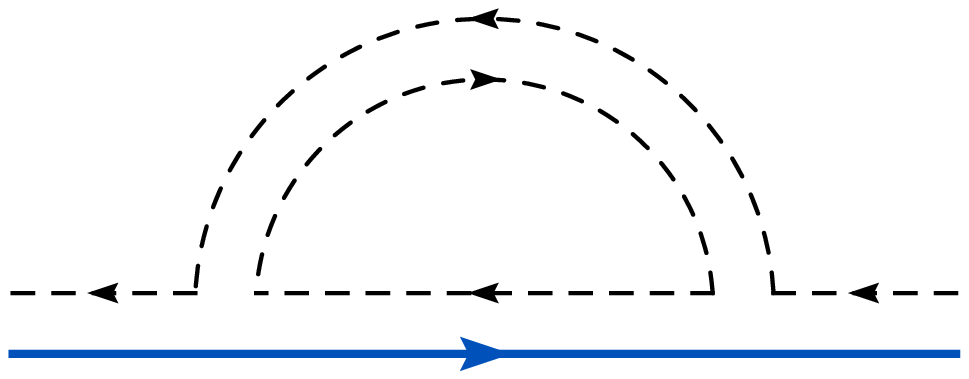} & $\;\;+\;\;$ & \epsfxsize=1.75in \epsffile{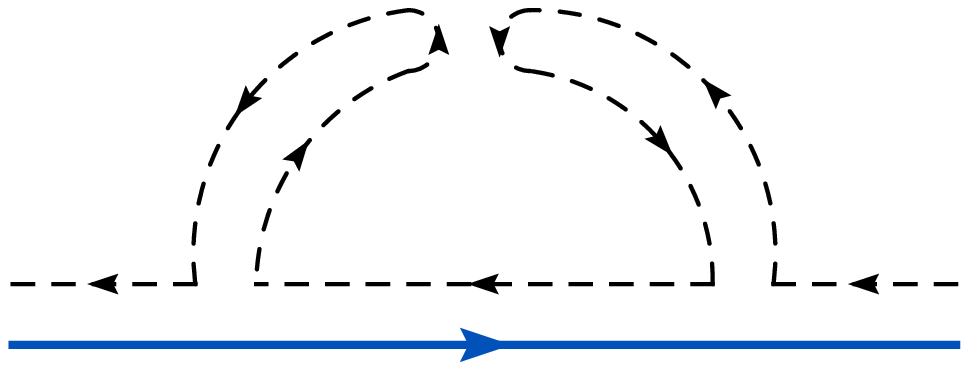} & $\;\;+\;\;$ & \epsfxsize=1.75in \epsffile{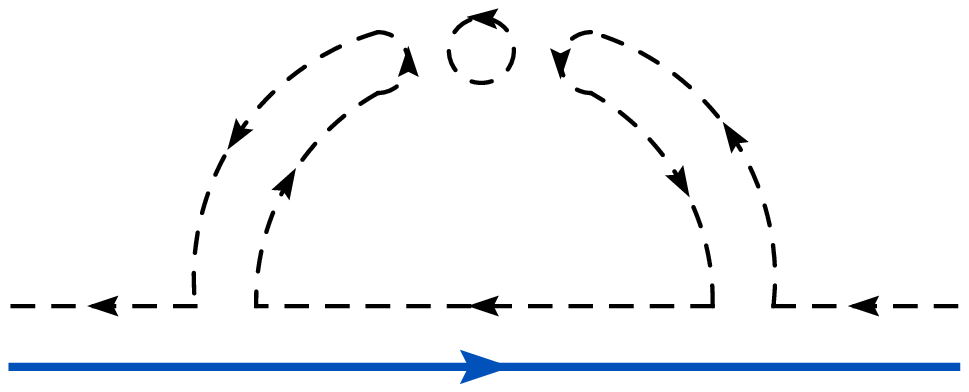} & \;\;\ldots\\
    & & (b) & &  \\
    \epsfxsize=1.75in \epsffile{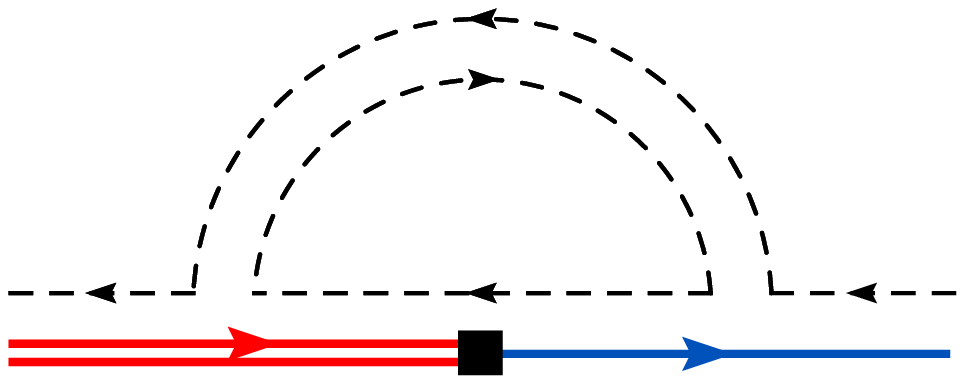} & $\;\;+\;\;$ & \epsfxsize=1.75in \epsffile{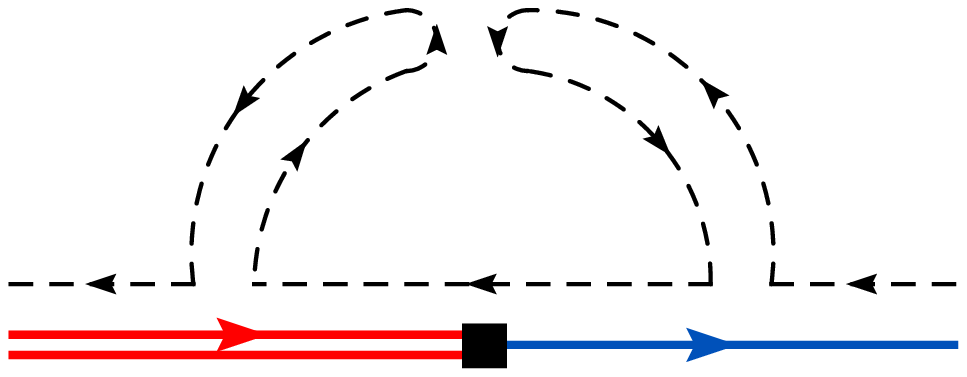} & $\;\;+\;\;$ & \epsfxsize=1.75in \epsffile{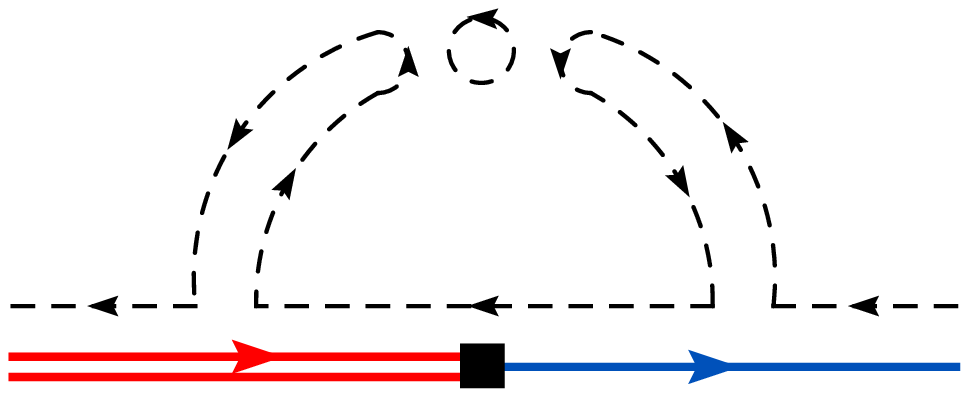} & \;\;\ldots \\
    & & (c) & &
    \end{tabular}
    \caption{Quark flow diagrams that contribute to $B\rightarrow D$ and $B\rightarrow D^*$.  The double line corresponds to the bottom quark within the $B$-meson, the single line corresponds to the charm quark within the $D^{(*)}$-meson, and the dashed lines correspond to staggered light quarks.  Diagram (a) renormalizes the $B$-meson wavefunction while (b) renormalizes the $D^{(*)}$-meson wavefunction.  Diagram (c) modifies the $B\rightarrow D^{(*)}$ vertex.}\label{fig:quark_flow}\end{figure}

The $B\to D$ and $B\to D^*$ matrix elements receive contributions
from the diagrams shown in Figure~\ref{fig:BtoD_Pion}.  It is
necessary, however, to consider these same diagrams at the quark
level in order to identify sea quark loops.  This is because, in
S$\chi$PT, all sea quark loops must be multiplied by $1/4$ in
order to describe data from staggered lattice simulations in which
the fourth-root of the quark determinant is taken to reduce the
number of tastes per flavor from 4 to 1.\footnote{Throughout this
paper we assume the validity of the $\sqrt[4]{\mbox{Det}}$ trick;
for a recent review of the status of the $\sqrt[4]{\mbox{Det}}$
trick see Ref.~\cite{Durr:2005ax}.}  Quark flow analysis also
allows identification of quark-disconnected hairpin diagrams,
which can only occur for taste $I$, $V$, and $A$ pion loops, that
have propagators with multiple poles.  At the quark level, two
vertices appear in the calculation of the $B\to D$ and $B\to D^*$
form factors;  they are shown in Figure~\ref{fig:vertices}.  The
$\overline{H}H\pi$ vertex comes from the LO heavy meson chiral
Lagrangian, Eq.~(\ref{eq:HMChPT}), and is proportional to $g_\pi$.
The $\overline{D}B$ vertex comes from the weak operator in
Eq.~(\ref{eq:BtoD_Op}).  Using these vertices,
Figure~\ref{fig:quark_flow} shows the same diagrams as in
Figure~\ref{fig:BtoD_Pion}, but at the level of quark flow.

\bigskip

We now present results for the form factors relevant for $B\rightarrow D$ and $B\rightarrow
D^*$ at zero recoil including taste-breaking effects due to the staggered light quarks.

For the 1+1+1 PQ theory, in which $m_u \neq m_d \neq m_s$:
\begin{eqnarray}\label{eq:h+comp}  h_{+}^{(B_x)PQ,1+1+1}(1)&=& 1 + \frac{X_{+}(\Lambda)}{m_c^2}+\frac{g^2_\pi}{48
\pi^2f^2}\Bigg\{\frac{1}{16}\sum_{\begin{subarray}{l}j=xu,xd,xs\\
\Xi=I,P,4V,4A,6T \end{subarray}} \!\!\!\!\!\!\! {F}_{j_\Xi}
+\frac{1}{3}\bigg[R^{[2,2]}_{X_I}\big(\{M^{(1)}_{X_I}\}
;\{\mu_I\}\big)\left(\frac{d{F}_{X_I}}{dm^2_{X_I}}\right)
\nonumber \\ && -\sum_{j \in
\{M^{(1)}_I\}}D^{[2,2]}_{j,X_I}\big(\{M^{(1)}_{X_I}\};\{\mu_I\}\big){F}_{j}
\bigg] +a^2\delta'_{V}\bigg[R^{[3,2]}_{X_V}\big(\{M^{(3)}_{X_V}\}
;\{\mu_V\}\big)\left(\frac{d{F}_{X_V}}{dm^2_{X_V}}\right) .
\nonumber
\\ &&
-\sum_{j \in \{M^{(3)}_V\}}D^{[3,2]}_{j,X_V}\big(\{M^{(3)}_{X_V}\};\{\mu_V\}\big){F}_{j}
 \bigg] +\big(V\rightarrow A\big)\Bigg\}, \end{eqnarray}
 \begin{eqnarray}\label{eq:hA1comp}  h_{A_1}^{(B_x)PQ,1+1+1}(1)&=& 1 + \frac{X_{A}(\Lambda)}{m_c^2}+\frac{g^2_\pi}{48
\pi^2f^2}\Big\{\frac{1}{16}\sum_{\begin{subarray}{l}j=xu,xd,xs\\
\Xi=I,P,4V,4A,6T \end{subarray}} \!\!\!\!\!\!\!
\overline{F}_{j_\Xi}
+\frac{1}{3}\bigg[R^{[2,2]}_{X_I}\big(\{M^{(1)}_{X_I}\}
;\{\mu_I\}\big)\left(\frac{d\overline{F}_{X_I}}{dm^2_{X_I}}\right)
\nonumber \\ && -\sum_{j \in
\{M^{(1)}_I\}}D^{[2,2]}_{j,X_I}(\{M^{(1)}_{X_I}\};\{\mu_I\})\overline{F}_{j}
\bigg] +a^2\delta'_{V}\bigg[R^{[3,2]}_{X_V}\big(\{M^{(3)}_{X_V}\}
;\{\mu_V\}\big)\left(\frac{d\overline{F}_{X_V}}{dm^2_{X_V}}\right)
\nonumber
\\ &&
-\sum_{j \in \{M^{(3)}_V\}}D^{[3,2]}_{j,X_V}\big(\{M^{(3)}_{X_V}\};\{\mu_V\}\big)\overline{F}_{j}
 \bigg] +\big(V\rightarrow A\big)\Bigg\}\,, \end{eqnarray}
where $x$ labels the light valence quark within the decaying $B_x$
meson. The first term inside the curly braces comes from diagrams
with sea quark loops; the flavor index $j$ runs over mesons made
of one valence quark and one sea quark and the taste index $\Xi$
runs over the sixteen pion tastes.
The residues $R^{[n,k]}_j$ and
$D^{[n,k]}_{j,l}$ are due to flavor-neutral hairpin propagators;
their explicit forms, along with the sets of masses $\{M_{X_\Xi}^{(i)}\}$, are given in the Appendix.
The second term in the curly braces (with coefficient $1/3$) comes from taste-singlet hairpins, while the third and fourth terms (with coefficients $a^2\delta'_V$ and $a^2 \delta'_A$) come from taste-vector and axial-vector hairpins, respectively.
The functions $F$ and
$\overline{F}$ are defined as
\bea
    F_j &\equiv& F\left(m_j,\Delta^{(c)}/m_j\right) \\
    \overline{F}_j &\equiv& F\left(m_j,-\Delta^{(c)}/m_j\right),
\eea
where
\begin{eqnarray}
F\left(m_j,x\right) &=& \frac{m_j^2}{x}\bigg\{x^3
\ln\frac{m_j^2}{\Lambda^2}+\frac{1}{3}x^3 -4x+2\pi \nonumber \\
&&
-\sqrt{x^2-1}(x^2+2)\left(\ln\Big[1-2x(x-\sqrt{x^2-1})\Big]-i\pi\right)\bigg\}
\nonumber
\\ && \longrightarrow (\Delta^{(c)})^2\ln\left(\frac{m_j^2}{\Lambda^2}\right)+{\cal O}[(\Delta^{(c)})^3]
\end{eqnarray}
and $m_j$ is the tree-level mass of a meson with flavor-taste
index $j$, given in Eq.~(\ref{eq:LOmass}). The analytic terms
proportional to $X_{+}(\Lambda)$ and $X_{A}(\Lambda)$ exactly
cancel the renormalization scale dependence of the $F$ terms.  It
is interesting to note that heavy-quark symmetry forbids the
presence of additional analytic terms such as those $\propto
\Str(\CM) = (m_u + m_d + m_s)$ or $\propto a^2$.  We have checked
that these expressions agree with the continuum partially quenched
ones when $a\to 0$ \cite{Savage:2001jw}.  In this limit, the
masses of all of the pion tastes become degenerate, so
$\frac{1}{16} \sum_\Xi F_{j_\Xi} \to F_j$.  Thus the sea quark
loop and taste-single hairpin contributions reduce to the
continuum PQ result, while the taste-vector and axial-vector
contributions, which are proportional to $a^2$, vanish.

For the 2+1 theory, in which $m_u=m_d \neq m_s$:
\begin{eqnarray}  h_{+}^{(B_x)PQ,2+1}(1)&=& 1 + \frac{X_{+}(\Lambda)}{m_c^2}+\frac{g^2_\pi}{16
\pi^2f^2}\Bigg\{\frac{1}{16}\sum_{\begin{subarray}{l}j=xu,xu,xs\\
\Xi=I,P,4V,4A,6T \end{subarray}} \!\!\!\!\!\!\!
F_{j_\Xi}+\frac{1}{3}\bigg[
R^{[2,2]}_{X_I}\big(\{M^{(5)}_{X_I}\};\{\mu_I
\}\big)\left(\frac{dF_{X_I}}{dm^2_{X_I}}\right) \nonumber \\ &&
-\sum_{j \in
\{M^{(5)}_I\}}D^{[2,2]}_{j_I,X_I}\big(\{M^{(5)}_{X_I}\};\{\mu_I\}\big)F_{X_I}
\Big] +a^2\delta'_{V}\bigg[R^{[3,2]}_{X_V}\big(\{M^{(7)}_{X_V}
\};\{\mu_V\}\big)\left(\frac{dF_{X_V},}{dm^2_{X_V}}\right)
\nonumber
\\ &&
-\sum_{j \in \{M^{(7)}_V\}}D^{[3,2]}_{j_V,X_V}\big(\{M^{(7)}_{X_V}\};\{\mu_V\}\big)F_{X_V},
 \bigg] +\big(V\rightarrow A\big)\Bigg\}, \end{eqnarray}
 \begin{eqnarray}  h_{A_1}^{(B_x)PQ,2+1}(1)&=& 1 + \frac{X_{A}(\Lambda)}{m_c^2}+\frac{g^2_\pi}{48
\pi^2f^2}\Bigg\{\frac{1}{16} \sum_{\begin{subarray}{l}j=xu,xu,xs\\ \Xi=I,P,4V,4A,6T \end{subarray}} \!\!\!\!\!\!\! \overline{F}_{j_\Xi} \nonumber\\
 &+& \frac{1}{3}\bigg[R^{[2,2]}_{X_I}\big(\{M^{(5)}_{X_I}\}
;\{\mu_I\}\big)\left(\frac{d\overline{F}_{X_I}}{dm^2_{X_I}}\right)
-\sum_{j \in \{M^{(5)}_I\}}D^{[2,2]}_{j,X_I}\big(\{M^{(5)}_{X_I}\};\{\mu_I\}\big)\overline{F}_{j}
\bigg] \nonumber \\
  &+& a^2\delta'_{V}\bigg[R^{[3,2]}_{X_I}\big(\{M^{(7)}_{X_V}\}
;\{\mu_V\}\big)\left(\frac{d\overline{F}_{X_V}}{dm^2_{X_V}}\right)
-\sum_{j \in \{M^{(7)}_V\}}D^{[3,2]}_{j,X_V}\big(\{M^{(7)}_{X_V}\};\{\mu_V\}\big)\overline{F}_{j}
 \Big] \nonumber \\
 &+& \big(V\rightarrow A\big)\Bigg\}. \end{eqnarray}

In the case of full (2+1) QCD, the expressions for the residues
simplify because QCD is a physical (unitary) theory without double
poles:
\begin{eqnarray}\label{eq:h+full} h_{+}^{(B_u)QCD,2+1}(1)&=&
1+\frac{X_{+}(\Lambda)}{m_c^2}+\frac{g^2_\pi}{16\pi^2f^2}\left[\frac{1}{16}\sum_{\Xi}\big(2F_{\pi_\Xi}+F_{K_\Xi}\big)
-\frac{1}{2}F_{\pi_I}+\frac{1}{6}F_{\eta_I} \right.\nonumber
\\ &&
+a^2\delta'_V\left(\frac{m^2_{S_V}-m^2_{\pi_V}}{(m^2_{\eta_V}-m^2_{\pi_V})(m^2_{\pi_V}-m^2_{\eta'_V})}F_{\pi_V}
+\frac{m^2_{\eta_V}-m^2_{S_V}}{(m^2_{\eta_V}-m^2_{\eta'_V})(m^2_{\eta_V}-m^2_{\pi_V})}F_{\eta_V}
\right. \nonumber \\ && \left. \left.
+\frac{m^2_{S_V}-m^2_{\eta'_V}}{(m^2_{\eta_V}-m^2_{\eta'_V})(m^2_{\eta'_V}-m^2_{\pi_V})}F_{\eta'_V}\right)
+\big(V\rightarrow A\big) \right], \end{eqnarray}
\begin{eqnarray}\label{eq:hA1full} h_{A_1}^{(B_u)QCD,2+1}(1)&=&
1+\frac{X_{A}(\Lambda)}{m_c^2}+\frac{g^2_\pi}{48\pi^2f^2}\left[\frac{1}{16}\sum_{\Xi}\big(2\overline{F}_{\pi_\Xi}+\overline{F}_{K_\Xi}\big)
-\frac{1}{2}\overline{F}_{\pi_I}+\frac{1}{6}\overline{F}_{\eta_I}
\right.\nonumber \\ &&
+a^2\delta'_V\left(\frac{m^2_{S_V}-m^2_{\pi_V}}{(m^2_{\eta_V}-m^2_{\pi_V})(m^2_{\pi_V}-m^2_{\eta'_V})}\overline{F}_{\pi_V}
+\frac{m^2_{\eta_V}-m^2_{S_V}}{(m^2_{\eta_V}-m^2_{\eta'_V})(m^2_{\eta_V}-m^2_{\pi_V})}\overline{F}_{\eta_V}
\right. \nonumber \\ && \left. \left.
+\frac{m^2_{S_V}-m^2_{\eta'_V}}{(m^2_{\eta_V}-m^2_{\eta'_V})(m^2_{\eta'_V}-m^2_{\pi_V})}\overline{F}_{\eta'_V}\right)
+\big(V\rightarrow A\big) \right]. \end{eqnarray}
\begin{eqnarray} h_{+}^{(B_s)QCD,2+1}(1)&=&
1+\frac{X_{+}(\Lambda)}{m_c^2}+\frac{g^2_\pi}{16\pi^2f^2}\left[\frac{1}{16}\sum_{\Xi}\big(F_{S_\Xi}+2F_{K_\Xi}\big)
-F_{S_I}+\frac{2}{3}F_{\eta_I} \right.\nonumber
\\ &&
+a^2\delta'_V\left(\frac{m^2_{S_V}-m^2_{\pi_V}}{(m^2_{S_V}-m^2_{\eta_V})(m^2_{S_V}-m^2_{\eta'_V})}F_{S_V}
+\frac{m^2_{\eta_V}-m^2_{\pi_V}}{(m^2_{\eta_V}-m^2_{S_V})(m^2_{\eta_V}-m^2_{\eta'_V})}F_{\eta_V}
\right. \nonumber \\ && \left. \left.
+\frac{m^2_{\eta'_V}-m^2_{\pi_V}}{(m^2_{\eta'_V}-m^2_{S_V})(m^2_{\eta'_V}-m^2_{\eta_V})}F_{\eta'_V}\right)
+\big(V\rightarrow A\big) \right], \end{eqnarray}
\begin{eqnarray} h_{A_1}^{(B_s)QCD,2+1}(1)&=&
1+\frac{X_{A_1}(\Lambda)}{m_c^2}+\frac{g^2_\pi}{48\pi^2f^2}\left[\frac{1}{16}\sum_{\Xi}\big(\overline{F}_{S_\Xi}+2\overline{F}_{K_\Xi}\big)
-\overline{F}_{S_I}+\frac{2}{3}\overline{F}_{\eta_I} \right.\nonumber
\\ &&
+a^2\delta'_V\left(\frac{m^2_{S_V}-m^2_{\pi_V}}{(m^2_{S_V}-m^2_{\eta_V})(m^2_{S_V}-m^2_{\eta'_V})}\overline{F}_{S_V}
+\frac{m^2_{\eta_V}-m^2_{\pi_V}}{(m^2_{\eta_V}-m^2_{S_V})(m^2_{\eta_V}-m^2_{\eta'_V})}\overline{F}_{\eta_V}
\right. \nonumber \\ && \left. \left.
+\frac{m^2_{\eta'_V}-m^2_{\pi_V}}{(m^2_{\eta'_V}-m^2_{S_V})(m^2_{\eta'_V}-m^2_{\eta_V})}\overline{F}_{\eta'_V}\right)
+\big(V\rightarrow A\big) \right], \end{eqnarray}
Note that there are separate formulae
for $B_{u,d} \rightarrow D^{(*)}_{u,d}$ and $B_s \rightarrow D^{(*)}_s$ in full QCD. This is in contrast to the PQ
expressions, which are valid for any choice of light quark flavor.
These results agree with the continuum full QCD ones when $a\to 0$
\cite{Randall:1993qg,Savage:2001jw}.

\section{Numerical Illustration of the $B\to D^*$ Form Factor}
\label{sec:num}

In this section we present a realistic picture of the behavior of
staggered lattice data for $B\to D^*$ and compare our S$\chi$PT
expression to actual staggered lattice data for $B\to D$.

\bigskip

\begin{figure}
\begin{center}
\includegraphics[scale=.45]{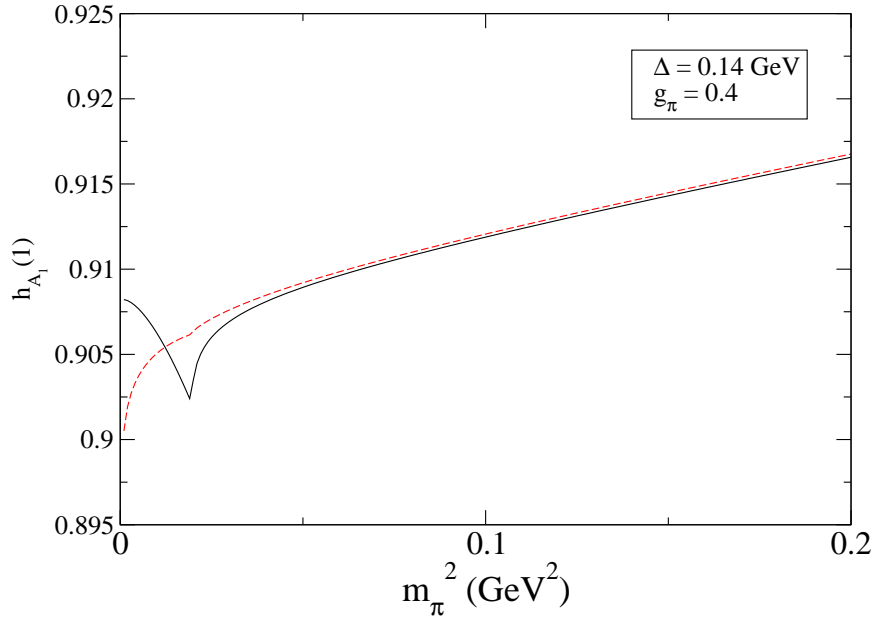}
\end{center}
\caption{Qualitative behavior of $h_{A_1}(1)$ vs. $m_\pi^2$.  The
overall linear contribution comes from matching to existing
quenched data \cite{Hashimoto:2001nb}. The curve with the large
cusp is the continuum expression, whereas the (dashed) curve with
the mild cusp includes staggered discretization effects. We use
the measured values of the pion mass-splittings and taste-breaking
hairpins from the MILC coarse lattices as input into the staggered
curve \cite{Aubin:2004fs}. \label{fit1}}
\end{figure}

Figure~\ref{fit1}, which shows the full QCD expression for
$h_{A_1}(1)$ vs. $m_\pi^2$, illustrates the importance of
accounting for staggered discretization errors in the
extrapolation of staggered lattice data.  There are currently no
unquenched staggered lattice data for $B\to D^*$ available, so we
have added a term linear in $m_\pi^2$ to the S$\chi$PT expression
for $h_{A_1}(1)$ and matched onto existing quenched data simulated
at heavy ($>500\MeV$) pion masses \cite{Hashimoto:2001nb}.  Thus
Figure~\ref{fit1} gives a realistic illustration of what the
chiral extrapolation of unquenched data for $h_{A_1}(1)$ from the
MILC coarse lattices ($a=0.125$ fm) might look like.  The
continuum expression for $h_{A_1}(1)$ has a characteristic cusp at
$m_\pi=\Delta_c$ where the internal $D$ goes on-shell.  The
staggered expression has a cusp in the same location due to the
taste pseudoscalar pion, which receives no taste-breaking shifts
to its mass, but the cusp is much milder.

It is worthwhile to discuss in some detail why the staggered cusp
is so mild, or, equivalently, how the staggered curve in
Figure~\ref{fit1} becomes the continuum curve when $a \rightarrow
0$.  A cusp occurs in $h_{A_1}$ every time the internal pion and
$D$ go on-shell in the $B\to D^*$ diagram.  For a staggered pion
of taste $\Xi$,  this happens when $m_\pi^2 + a^2 \Delta_\Xi =
\Delta_c^2$, where $m_\pi^2$ is the tree-level mass of the lattice
Goldstone pion and $a^2 \Delta_\Xi$ is the taste-breaking mass
correction.  Thus, in Figure~\ref{fit1}, there is a cusp in the
staggered curve every time $m_\pi^2 = \Delta_c^2 - a^2\Delta_\Xi$.
On the MILC coarse lattices, all of the $\CO(a^2)$ mass-splittings
are greater than $\Delta_c=0.14 \GeV$, so the additional heavy
staggered pions do not produce cusps in Figure~\ref{fit1}.  The
single staggered cusp due to the lattice Goldstone pion is small
because it is weighted by $1/16$ (from the average over pion
tastes in the loop) as compared to the continuum one.  As the
lattice spacing is reduced, more and more tastes will be able to
produce cusps to the left of the continuum one.  These cusps will
begin at $m_\pi^2=0$ and move to the right as the lattice spacing
becomes smaller.  Finally, at $a=0$, all of the cusps from the
non-Goldstone pions will come to rest at the location of the
continuum one, and the sum of these sixteen staggered cusps will
equal the single cusp in the continuum curve. In addition to
softening the cusp, the heavy staggered pions  decrease the
curvature due to chiral logarithms in $h_{A_1}$;  this is a
generic effect of taste-breaking. Thus the staggered data are
expected to be almost linear, even when the continuum result is
not.

In practice, one extrapolates staggered lattice data to the
continuum by first fitting to Eq.~(\ref{eq:hA1full}) and then
removing taste-breaking discretization errors by setting the terms
proportional to $a^2$ in Eq.~(\ref{eq:hA1full}) to zero.   We note
that simulations are not likely to be sensitive to the cusp
anytime soon, even if staggering did not smooth it out, because
the cusp only occurs at values very close to the physical pion
mass.  Thus, in the case of $B\to D^*$, it is especially important
to use S$\chi$PT to extrapolate to the physical light quark
masses.

\begin{figure}
\begin{center}
\includegraphics[scale=.45]{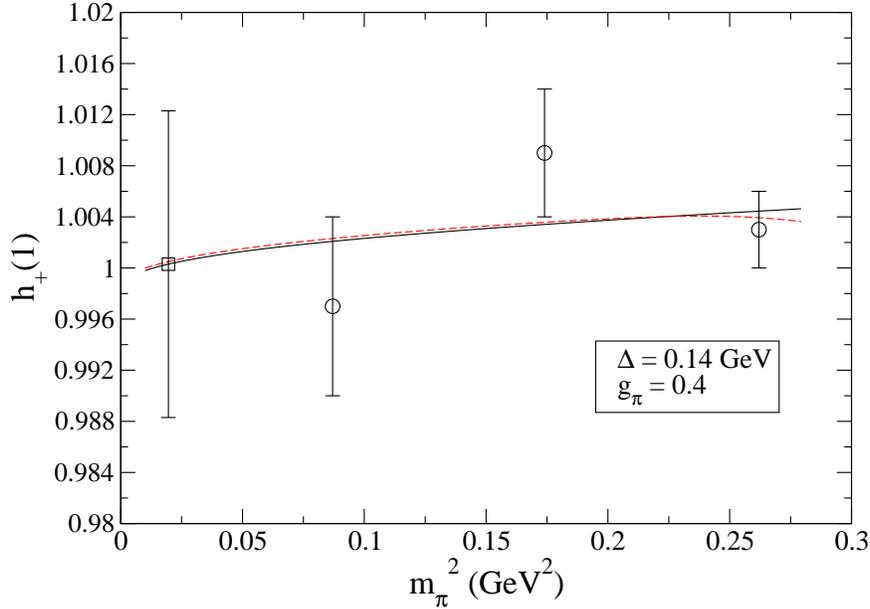}
\end{center}
\caption{$h_+(1)$ vs. $m_\pi^2$.   The three full QCD data points
(circles) were calculated on the MILC coarse lattices ($a=0.125$
fm) \cite{Okamoto:2004xg}.  The upper (dashed) curve is a fit to
the data using the complete staggered formula, while the lower
(solid) curve is the continuum extrapolated curve. The square is
the extrapolated value of $h_+(1)$ at the physical pion mass with
error bars. \label{fit2}}
\end{figure}

We can also directly apply our S$\chi$PT expression to the
available unquenched data for $B\to D$ \cite{Okamoto:2004xg}.
Figure~\ref{fit2} shows $h_+(1)$ vs. $m_\pi^2$ in full QCD.  In
this case, the pion and $D^*$ in the loops of the diagrams of
Figure~\ref{fig:BtoD_Pion} cannot go on-shell, so there is no cusp
as in the case of $B\to D^*$. The dashed line is the result of a
fit of the staggered expression to the three data points. The
solid line is the continuum extrapolated curve, while the square
is the continuum extrapolated value of $h_+(1)$ at the physical
value of the pion mass, with error bars. The difference between
these curves is very small, and the extrapolated value hardly
differs from the result of a naive linear fit. Nevertheless, the
S$\chi$PT analysis is useful for this quantity because it
demonstrates that the systematic errors associated with the chiral
extrapolation are small.

\section{Finite Volume Effects in $B\to D^{(*)}$}
\label{sec:FVE}

The functions $F$ and $\bar{F}$, which appear in $h_+(1)$ and
$h_{A_1}(1)$, respectively, are modified by the finite spatial
extent of the lattice. Using the formulae for finite volume
corrections to typical HM$\chi$PT integrals given in
Ref.~\cite{Arndt:2004bg}, we find that $F(m,\Delta)$ receives the
following correction due to the finite lattice volume:
\bea\label{eq:fv} \delta F_{FV}(m,\Delta,\textrm{L}) & = &
\sum_{\vec{n}\neq\vec{0}} \left( \frac{m^2}{128 x y^3}\right)
e^{-y}\Bigg\{\pi e^{x^2 y/2} \bigg[y^5x^{12}-(y-16)y^4x^{10}
\nonumber
\\ &&  +2 (y+24)y^3x^8-16(y-2)y^3x^6+96 y^3x^4-128 y^3 x^2+256
y^2\bigg] \nonumber \\ && +\sqrt{2\pi
y}\bigg[-y^4x^{11}+(y-15)y^3x^9-(3y+35)y^2x^7 \nonumber
\\ &&  +(16y^2-27y+15)yx^5 +(-112y^2+11y-9)x^3+256y^2x \bigg]
\nonumber \\ && -256\pi y^2 -\pi y^2 e^{x^2
y/2}\textrm{erf}\left(\frac{x\sqrt{y}}{\sqrt{2}}\right) \bigg[
y^3x^{12}-(y-16)y^2x^{10}  \nonumber
\\ &&  +2(y+24)yx^8-16(y-2)yx^6+96yx^4-128yx^2+256 \bigg]
\Bigg\},
   \eea
where $x=\Delta/m$ as before, $y=nm\textrm{L}$, and
$n=\sqrt{\vec{n}^2}$.  The correction to $\bar{F}$ is identical
except for $x\to -x$.  This formula was derived as a series
expansion in $1/(nm\textrm{L})$.  In our numerical evaluation of
this formula we truncate the sum to the values of $n=1$,
$\sqrt{2}$, $\sqrt{3}$, $\sqrt{4}$, $\sqrt{5}$ and
$\sqrt{6}$.\footnote{Ref.~\cite{Arndt:2004bg} determined that
truncating the sum at $n=\sqrt{5}$ approximates the full answer
well ($\sim 3\%$) for $mL \geq 2.5$.} An expansion in $x=\Delta/m$
shows that the leading contribution to $\delta F_{FV}$ is
proportional to $\Delta^2$, as expected:
\bea \delta F_{FV}(m,\Delta,\textrm{L}) & = & \sum_{\vec{n}\neq\vec{0}}
\sqrt{\frac{\pi}{2}}\left(\frac{m^2x^2}{192
y^{5/2}}\right)e^{-y}\left(128y^3-336y^2+33y-27\right) + {\cal
O}(x^3). \eea

\begin{figure}
\begin{center}
\includegraphics[scale=.45]{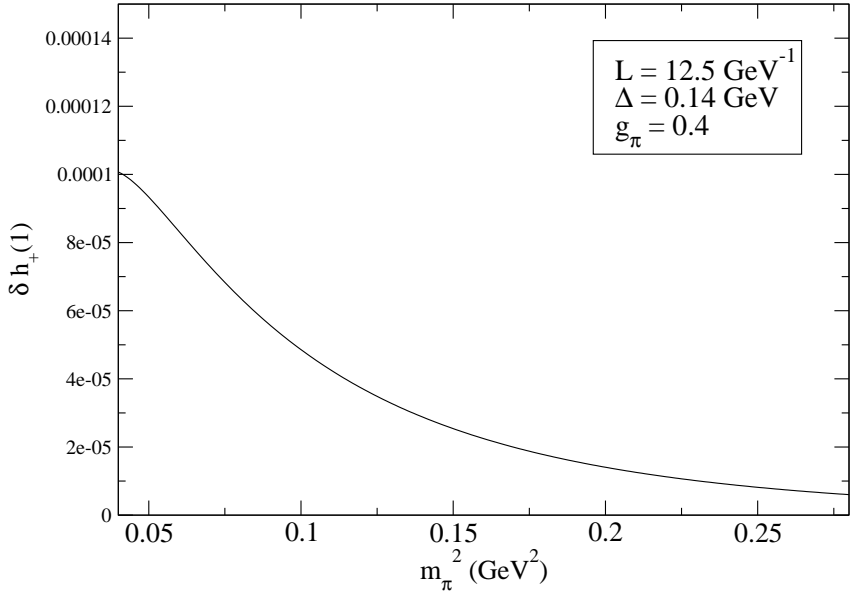}
\end{center}
\caption{Finite volume correction to $h_+(1)$ as a function of
$m_\pi^2$.  Recall that $h_+(1)$ is close to 1, so these
corrections are smaller than one part in $10^{-4}$ in current
staggered simulations. \label{fv}}
\end{figure}

Figure~\ref{fv} shows the contribution to $h_+(1)$ in full QCD
from finite volume effects for the MILC coarse lattice ($a=0.125$
fm, $\textrm{L}=2.5$ fm).  Recall from Figure~\ref{fit2} that
$h_+(1)$ is close to one, whereas the finite volume corrections in
Figure~\ref{fv} are less than $10^{-4}$ in the range of pion
masses relevant for current staggered lattice simulations. The
size of finite volume corrections to $h_{A_1}(1)$ are similarly
small. We therefore conclude that finite volume errors are
negligible in both the $B\to D$ and $B\to D^*$ form factors, and
can be accounted for as an overall systematic error in lattice
calculations, rather than subtracted before the chiral
extrapolation.

\section{Conclusions}
\label{sec:Conc}

In this work we have calculated the $B\to D$ and $B\to D^*$ form
factors at zero recoil to NLO in S$\chi$PT. We have presented
expressions for both a ``1+1+1'' partially quenched theory ($m_u
\neq m_d \neq m_s$) and a ``2+1'' partially quenched theory
($m_u=m_d \neq m_s$), as well as for full (2+1) QCD. These
formulae apply to simulations in which only the light quark is
staggered. They include $\CO(a^2)$ taste-breaking discretization
errors, and are necessary for correct continuum and chiral
extrapolation of staggered $B\to D^{(*)}$ lattice data. Use of
these expressions, along with the double ratio method of
Ref.~\cite{Hashimoto:2001nb} and in combination with experimental
input, should allow a precise determination of the CKM matrix
element $|V_{cb}|$.

\section*{Acknowledgments}

We thank Masataka Okamoto for sharing his $h_+(1)$ lattice data,
David Lin and Steve Sharpe for useful discussions, and Andreas
Kronfeld and Paul Mackenzie for reading the manuscript.  This
research was supported by the DOE under grant no.
DE-AC02-76CH03000.

\appendix
\section*{\bf APPENDIX}
\setcounter{equation}{0} \setcounter{section}{1}
\renewcommand{\theequation}{A\arabic{equation}}

In this section we collect formulae necessary for understanding
our form factor results.  We follow the notation of
Ref.~\cite{Aubin:2003uc}.

\bigskip

The residues $R^{[n,k]}_j$ and $D^{[n,k]}_{j, l}$ appear because
of single and double poles, respectively, in the flavor-neutral
hairpin propagators:
\begin{eqnarray} R^{[n,k]}_j(\{m\},\{\mu\})  &\equiv & \frac{\prod_{a=1}^k
(\mu^2_a-m^2_j)}{\prod_{i\neq j} (m^2_i-m^2_j)}, \nonumber \\
D^{[n,k]}_{j, l}(\{m\},\{\mu\})  &\equiv &
-\frac{d}{dm^2_l}R^{[n,k]}_j(\{m\},\{\mu\}). \end{eqnarray}

Once one takes the mass of the overall flavor-taste singlet pion
(which corresponds to the physical $\eta'$) to infinity, the
relationships among the taste-singlet pion masses simplify:
\begin{eqnarray} m^2_{\pi_I}  &= & m^2_{U_I} = m^2_{D_I}, \nonumber \\
m^2_{\eta_I} & = & \frac{m^2_{U_I}}{3} + \frac{2m^2_{S_I}}{3}, \end{eqnarray}
Thus the following mass combinations appear in the 1+1+1 ($m_u \neq m_d \neq m_s$) PQ result:
\begin{eqnarray} \{M_X^{(1)}\} &\equiv & \{m_{\pi^0},m_\eta, m_X \}, \nonumber \\
    \{M_X^{(3)}\} &\equiv & \{m_{\pi^0},m_\eta, m_{\eta'}, m_X \}, \nonumber\\
    \{\mu\} &\equiv &\{m_U,m_D,m_S\} \end{eqnarray}

When the up and down quark masses are degenerate, the pion mass
eigenstates become:
\begin{eqnarray} m^2_{\pi_V}  & = & m^2_{U_V} = m^2_{D_V}, \nonumber \\
    m^2_{\eta_V}  & = & \frac{1}{2}\left(m^2_{U_V}+m^2_{S_V}+\frac{3}{4}a^2\delta'_V - Z
    \right),\nonumber \\
    m^2_{\eta'_V}  & = & \frac{1}{2}\left(m^2_{U_V}+m^2_{S_V}+\frac{3}{4}a^2\delta'_V + Z
    \right),\nonumber \\
    Z & \equiv & \sqrt{(m^2_{S_V}-m^2_{U_V})^2 - \frac{a^2\delta'_V}{2}(m^2_{S_V}-m^2_{U_V})+\frac{9(a^2\delta'_V)^2}{16}}, \end{eqnarray}
Thus the following mass combinations appear in the 2+1 ($m_u = m_d \neq m_s$) PQ result:

\begin{eqnarray} \{M_X^{(5)}\} &\equiv& \{m_\eta, m_X \}, \nonumber \\
    \{M_X^{(7)}\}&\equiv& \{m_\eta, m_{\eta'}, m_X \}, \nonumber \\
    \{\mu\}&\equiv&\{m_U,m_S\} \end{eqnarray}

\bibliography{BtoD_V2}

\end{document}